\documentclass[11pt,twoside]{article}
\usepackage{asp2014}

\aspSuppressVolSlug
\resetcounters

\begin{document}

\title{The Recent Developmental Status of SNEGRAF: a Web-Based Gravitational Wave Signal Analyzer}

\author{Satoshi~Eguchi$^1$, Shota~Shibagaki$^1$, Kazuhiro~Hayama$^1$, and Kei~Kotake$^1$}
\affil{$^1$Department of Applied Physics, Faculty of Science, Fukuoka University, Fukuoka, Japan; \email{satoshieguchi@fukuoka-u.ac.jp}}

\paperauthor{Satoshi~Eguchi}{}{0000-0003-2814-9336}{Fukuoka University}{Department of Applied Physics, Faculty of Science}{Fukuoka City}{Fukuoka Prefecture}{814-0180}{Japan}
\paperauthor{Shota~Shibagaki}{shibagaki@fukuoka-u.ac.jp}{}{Fukuoka University}{Department of Applied Physics, Faculty of Science}{Fukuoka City}{Fukuoka Prefecture}{814-0180}{Japan}
\paperauthor{Kazuhiro~Hayama}{hayama@fukuoka-u.ac.jp}{}{Fukuoka University}{Department of Applied Physics, Faculty of Science}{Fukuoka City}{Fukuoka Prefecture}{814-0180}{Japan}
\paperauthor{Kei~Kotake}{kkotake@fukuoka-u.ac.jp}{}{Fukuoka University}{Department of Applied Physics, Faculty of Science}{Fukuoka City}{Fukuoka Prefecture}{814-0180}{Japan}



  
\begin{abstract}
Unveiling physical processes in a supernova is one of challenging topics of modern physics
and astrophysics since that event is due to particle physics on a stellar scale and tightly
related to nucleosynthesis in Universe.
Multi-messenger astronomy, a combination, such as of electromagnetic-wave, gravitational-wave,
and neutrino observations, will be a breakthrough to the puzzle.
To boost the research, we released a web-based gravitational wave signal analyzer
``SuperNova Event Gravitational-wave-display in Fukuoka (SNEGRAF)'' last year \citep{2019ASPC..523..493E}.
We are now working on an integration of the application with the RIDGE pipeline,
which is for a coherent network analysis between the LIGO, VIRGO, and KAGRA observations
\citep{2007CQGra..24S.681H}, and implemented in MATLAB.
In the basic design phase, we decided to wrap RIDGE with a simple Python script and make it
listen for connections from SNEGRAF.
This design enables these two programs to be hosted on different servers independently,
and minimizes the cyber risks of RIDGE.
In this paper, we report the current developmental status of our system.
\end{abstract}

\section{Introduction}

A supernova explosion is one of the most energetic phenomena
in Universe, and it takes place in the last stage of
stellar evolution.
In Fukuoka University, we have a special team
aimed to reveal the physical mechanisms of core-collapse supernovae (CCSNe)
from both sides of theories (numerical simulations)
and observations.
The central engine of a CCSN is in an extremely condensed
state and photons can hardly escape, hence new observational methods
utilizing neutrinos and gravitational waves, to which
such dense materials are even transparent, are crucial;
these observations combined with classical electromagnetic
ones are referred to as multi-messenger astronomy.

When compared to events of neutron star mergers and
black hole ones, our understanding of the gravitational waves
radiated during a CCSN is rather limited.
Coherent network analysis (CNA) technique,
which coordinates individual observations by different worldwide
gravitational-wave detectors, provides us adequate information
even about a transient source including a CCSN since the method
can reduce the impact of temporal and unmodelled detector
noises on the data \citep{2006PhRvD..74h2005C}.
RIDGE is an implementation of CNA algorithms, and its robustness
has been confirmed by intensive reviews from the aspect of ``physics''
\citep{2007CQGra..24S.681H,2015PhRvD..92l2001H}.

\section{SNEGRAF}

``SuperNova Event Gravitational-wave-display in Fukuoka (SNEGRAF)''
is a web application to analyze gravitational-wave signals provided
by our team in order to boost the astrophysics of supernovae,
and was presented in ADASS2018 \citep{2019ASPC..523..493E}.
Although SNEGRAF at that time could just display FFT results and
the signal-to-noise ratio together with the analytic sensitivity
curve of KAGRA from the users' inputs,
our intensive improvements of the software made over the past year, for
an integration of SNEGRAF with the RIDGE pipeline,
allows the users to perform much more realistic simulations of
gravitational-wave detections with LIGO, Virgo, and KAGRA
based on their model waveforms, which are given as
a character-separated-values (CSV) file consisting of
a list of $\left(t, h_{+} \left( t \right) , h_{\times} \left( t \right) \right)$,
where $t$, $h_{+} \left( t \right)$, and $h_{\times} \left( t \right)$
are time in units of seconds, the plus and cross modes of
the gravitational waves at the moment $t$, respectively.
We are now in the final phase of unit testing for
both software;
we will be able to make the brand-new SNEGRAF public by the end of 2019.

\begin{figure}
 \plotone{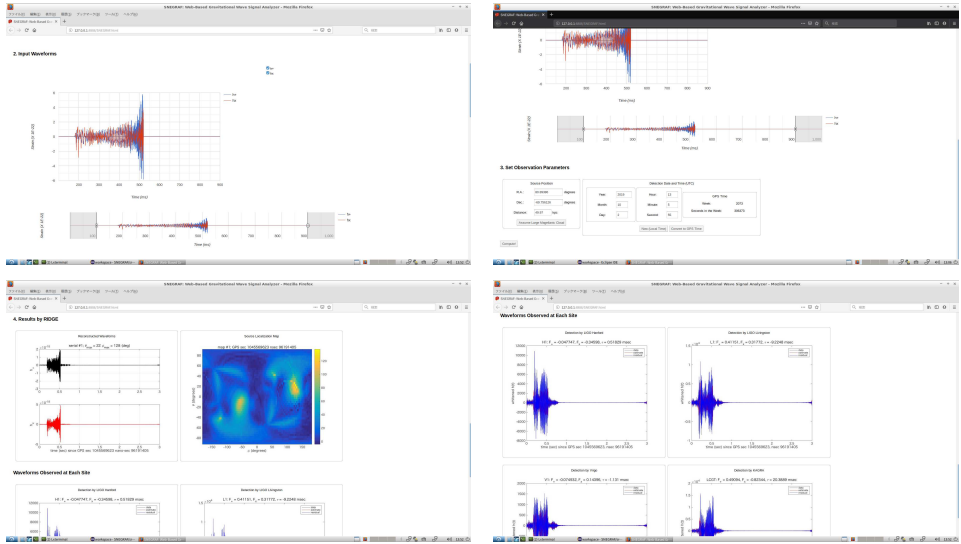}
 \caption{Screenshots of SNEGRAF.
  From left to right and top to bottom, the input waveform viewer,
  source parameter dialog box, reconstructed waveforms and source
  localization sky map by RIDGE, and ``observed'' waveforms
  at each site, respectively.}
 \label{P2-9_f1}
\end{figure}

Figure~\ref{P2-9_f1} shows screenshots of the latest development
version of SNEGRAF.
Once a user uploads his/her theoretical waveforms
(the top left panel in Figure~\ref{P2-9_f1}),
a panel asking the ``source'' information (right ascension, declination,
and the distance to the source) and ``detection'' date and time
appears (the top right panel in Figure~\ref{P2-9_f1}).
While RIDGE internally uses a GPS time representation,
we made SNEGRAF accept Coordinated Universal Time (UTC)
with consideration of leap seconds,
since UTC is the standard in ``traditional'' (or electromagnetic) astronomy;
the user can easily check the feasibility of follow-up observations especially
with X-ray and gamma-ray satellites, which have different observation windows
depending on a source direction, in case that such an event should really occur.

A click on ``Compute!'' button invokes RIDGE with the above parameters
together with the uploaded waveforms.
In the process, RIDGE convolutes the waveforms with the response and idealized
Gaussian noises of each detector,
performs a waveform reconstruction, and makes a source localization sky map.
Then these results including ``virtually observed'' signals by respective
detectors are sent back to SNEGRAF as scalable-vector-graphics (SVG) files;
SNEGRAF presents them as static images on the screen
(the bottom panels in Figure~\ref{P2-9_f1}).

\section{System Design}

\begin{figure}
 \plotone{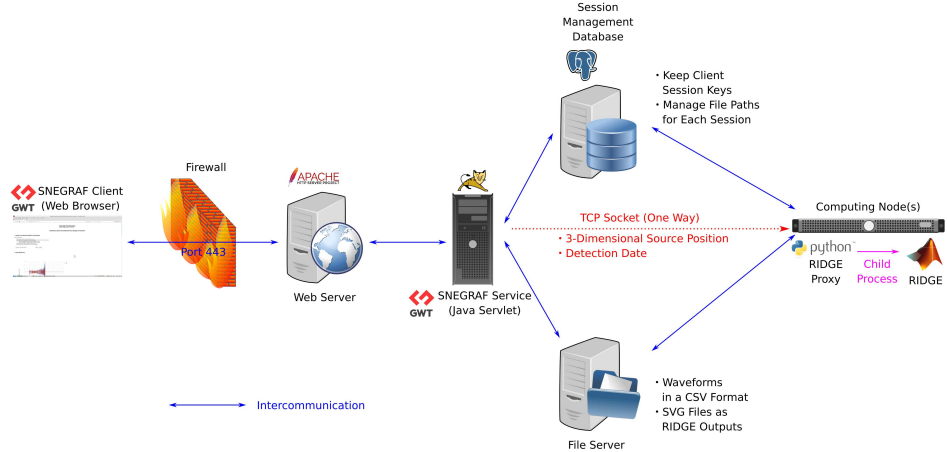}
 \caption{The schematic system diagram of SNEGRAF.
 The SNEGRAF client and server are implemented with
 GWT and communicate with each other over the HTTPS protocol.
 There is a lightweight proxy server written in Python
 which runs the RIDGE pipeline as its child processes.
 The SNEGRAF server and proxy server can exchange
 data just through a one-way TCP socket
 (from the servlet to the proxy) and a file
 server.
 There is a database server to manage sessions
 and their related files.
} \label{P2-9_f2}
\end{figure}

Figure~\ref{P2-9_f2} shows a diagram of our system.
For rapid development and maximum utilization of existing software
resources, we adopted GWT (previously known as Google Web Toolkit)
for a framework of SNEGRAF \citep{2019ASPC..523..493E};
GWT generates both sever-side (Java servlet) and client-side
(JavaScript) codes from one Java source file.

The RIDGE pipeline is written in MATLAB.
From the viewpoint of software engineering, RIDGE has not been
in long-term continuous operation as a software service.
In addition, a MATLAB application always runs on MATLAB kernel
even when it is packaged by MATLAB Compiler;
we have to be aware of risks similar to SQL injection attacks against
the kernel if we make RIDGE public to the Internet directly.
To these concerns, we took a very simple approach:
an isolation of RIDGE from both the Internet and a web server
at hardware level.

An application server running the SNEGRAF servlet and a computing
node running RIDGE can communicate with each other just through
a file server, which stores the users' inputs as character-separated-values
(CSV) files and SVG files produced by RIDGE, and
through a one-way transmission-control-protocol (TCP) socket
to send source parameters as CSV strings,
from the application server to the computing node.
Elsewhere than the file server, any files exist just as
Base64 encoded strings.
Since we would not like to make any changes to RIDGE
for awaiting a connection from the servlet, we implemented
a simple proxy server which invokes RIDGE as a child process
in a worker thread in Python, and run it on the computing node.
This design brings us a favorable by-product;
with another few tens of lines for a simple round-robin
scheduler in SNEGRAF, we can easily scale the system out
just by adding computing nodes.

\section{By-Products of Our Work}

\subsection{Time Utilities}

RIDGE internally uses a GPS time representation, which was defined
as the Coordinated Universal Time (UTC) at 00:00:00 on 1980 January 6,
behind 19 seconds against the International Atomic Time (TAI) at the moment,
and is synchronized with TAI since then.
Note that there is no leap second in GPS time.
Since we require a nanosecond precision of time with consideration of
leap seconds, we implemented time utility classes for GWT environments
from scratch.
While the most existing libraries use Julian Days (JDs) internally,
we adopted Modified Julian Days (MJDs) represented in units of seconds
as a 64-bit integer for accuracy.

\subsection{Pull-Request to GWT Charts}

The interactive chart panel to display uploaded
waveforms is implemented with GWT Charts,
a class library of Google Charts for GWT environments.
GWT Charts seems not to be maintained for a while;
the API loader interface for Google Charts changed
long, long time ago, but it was not reflected in GWT Charts.
On 2019 January 1, GWT Charts temporally stopped
working by a maintenance by Google, possibly due to
disablement of old loader codes.
We completely rewrote the core classes of GWT Charts to 
load the Google Charts modules, and made a pull request
to the upstream GitHub repository\footnote{\url{https://github.com/satoshieguchi/gwt-charts/tree/new_api_loader_for_pullreq}}.

\acknowledgements

This work is partially supported by JSPS KAKENHI Grant Number JP19K12244,
JP19K03896, JP17H06364, JP17H01130, JP17H06357
and by the Central Research Institute of Stellar Explosive Phenomena
(REISEP) of Fukuoka University
and the associated research projects (Nos.171042, 177103).

\bibliography{P2-9}


\end{document}